\documentstyle[pra,aps,epsfig]{revtex}
\begin{document}
\draft
\title{A PROPOSED INTERPRETATION OF MICROWAVE IONIZATION OF RYDBERG ALKALI ATOMS}
\author{Luca Perotti}
\address{Centro per lo studio dei sistemi dinamici, Universit\'a degli studi dell'Insubria, Como}
\date{\today}
\maketitle

\begin{abstract}
{\bf Abstract:} I show that the theory developed for quantum delocalization of an excited Hydrogen atom in a two-frequency linearly polarized microwave field \cite{cas} can be used with a few adaptions to explain the quantum delocalization of Alkali atoms in a linearly polarized monochromatic field in the regime where it deviates from Hydrogenic behaviour. Comparison with numerical \cite{buc,buc1,krug} and laboratory \cite{arn} experiments is satisfactory: apart from a constant factor, independent from the microwave frequency, the same dependence from the initial quantum number is observed.
\end{abstract}


\section{Introduction}

Recent numerical studies \cite{buc,buc1,krug} have shown that the ionization threshold of Rydberg Alkali atoms in a linearly polarized monochromatic microwave field with frequency $\omega$ agrees well with that of Hydogen atoms for high initial quantum number $n_0$ ($n_0\geq \omega^{-1/3}$) but strongly deviates from it for low $n_0$.

This deviation is attributed to the reduced spacing of levels due to the breaking of the angular momentum degeneracy by the core electrons: intuitively, the presence of low angular momentum levels in between the rungs of the high angular momentum Hydrogen-like ladder allows the electron an efficient ionization path even when the field frequency is lower than the nearest level transition frequency of Hydrogen.

No evaluation of the threshold in this regime has yet been given. Direct application of the Hydrogen quantum delocalization formula from Ref. \cite{chi} requires the assumption that the details of the underlying classical dynamics are not important as long as it is globally chaotic. This is not the case: when $n_0$ becomes lower than the value at which the Hydrogen classical chaos theshold crosses the quantum delocalization threshold ($n_0\cong \omega^{-1/3}$), the Alkali simulations \cite{buc1} show an abrupt change in the $n_0$-dependence of the ionization field intensity threshold from the Hydrogen-like inverse one to a flat one. I propose an evaluation in terms of a local correspondence between the biggest phase space structures of an Alkali atom in a linearly polarized monochromatic microwave field and those of a monodimensional Hydrogen atom in a linearly polarized {\it bichromatic} microwave field. In agreement with Ref. \cite{buc1} the threshold thus obtained does not depend on $n_0$.

This application of the theory of quantum delocalization of an excited Hydrogen atom in a two-frequency microwave field \cite{cas} also provides indirect experimental validation for it: there have been experimental works with bichromatic microwave fields \cite{due}, but they were out of the range of validity of the theory so that it has never been tested in practice.

\section{THE MODEL}

The motion of the perturbed system being restricted to a plane, I consider a two dimensional Alkali atom model \cite{co} in the $\{x,z\}= \{r,\theta\}$ plane:  
\begin{equation}
H_0 = {1\over 2}\left(p_r^2+{{p_{\theta}^2}\over{r^2}}\right) - {1\over r} - {{\beta e^{-\alpha r}}\over r},
\end{equation}
where the last term descibes the nonhydrogenic core potential \cite{no1}.
I then make a canonical transformation to the action angle variables $\{I,J,\lambda,\mu\}$ thus obtaining the Hamiltonian: 
\begin{equation}
H_0 = {1\over {2(I -\delta(J))^2}},
\end{equation}
where $I$ is the total action, $J$ is the angular momentum, and the ``quantum defect" $\delta(J)$ is a  monotonic decreasing function going to zero for $J\to I$; extended numerical tests have shown no discernible dependance of $\delta$ from $I$. 
The frequencies associated to the two actions are therefore:
\begin{equation}
\omega_I = {1\over {(I-\delta)^3}},\hspace{.5in}\omega_J = {{-{{d\delta}\over{dJ}}}\over {(I-\delta)^3}}= {{\left|{{d\delta}\over{dJ}}\right|}\over {(I-\delta)^3}}.
\end{equation}
 
To describe the atom-field interaction I now introduce the usual dipole approximation term $H_1 = Fz\cos\omega t = F r \cos\theta\cos\omega t$, where $F$ is the electric field amplitude and $\omega$ its frequency, and I Fourier expand it in $\lambda$ and $\mu$:
\begin{equation}
H_1 = F\Sigma_{n,m}V_{n,m}(I,J)\cos(n\lambda+m\mu-\omega t);
\end{equation}
where the sum extends to both positive and negative values of the indices $n$ and $m$, and the coefficients $V_{n,m}(I,J)$ are the semiclassical matrix elements for transitions with $\Delta I = n$ and $\Delta J = m$.

In extended phase space $\{I,J,K,\lambda,\mu,\psi\}$, with $K$ the ``photon number", the Hamiltonian therefore is:
\begin{equation}
H^{(Al)}=-{1\over {2(I -\delta(J))^2}}+\omega K +F\Sigma_{n,m}V_{n,m}(I,J)\cos(n\lambda+m\mu-\psi).
\label{Al}
\end{equation}

\section{CORRESPONDENCE}

The extended phase space Hamiltonian for a hydrogen atom in a two-frequency microwave field with frequencies $\omega_1$ and $\omega_2$ and field amplitudes $F_1$ and $F_2$ reads \cite{jens}:
\begin{equation}
H^{(2f)}=-{1\over {2I^2}}+\bar{\omega} K+ \Delta J +\Sigma_{n}V_{n}(I)[F_2\cos(n\lambda+ \mu-\psi)+F_1\cos(n\lambda- \mu-\psi)],
\label{2f}
\end{equation}
where $\bar{\omega}=(\omega_1+\omega_2)/2$ and $\Delta=(\omega_1-\omega_2)/2$ \cite{note}.

There are some similarities between eq. (\ref{2f}) and eq. (\ref{Al}); my aim is to find a suitable choice of the four parameters $F_1,F_2, \omega_1$, and $\omega_2$ so that the local phase space structure relevant to ionization is similar for the two Hamiltonians. A global correspondence is not possible as the two Hamiltonians are clearly non-equivalent: the functional dependence on $I$ and $J$ of the noninteracting part of the Hamiltonians is different in the two cases. Moreover, the atom-field interaction term in eq. (\ref{Al}) has more terms than the one in eq. (\ref{2f}) which only has the $m=\pm1$ ones; but, since these are those corresponding to the most important resonances (the dipole ones) in eq. (\ref{Al}), this is not a big problem.

From the amplitudes of the interaction terms we immediately have:
\begin{eqnarray}
F_1=F{{V_{n,-1} (I,J)}\over{V_{n'}(I')}}, \hspace{1.in} F_2=F{{V_{n,+1} (I,J)}\over{V_{n'}(I')}} ,
\end{eqnarray}
where, for the moment we leave undetermined whether $I$ equals $I'$ and $n$ equals $n'$.

More problematic is dealing with the atomic part for the determination of the frequencies: the $(n,\pm1)$ resonance condition reads
\begin{equation}
{n\over {(I_0-\delta(J_0))^3}}\pm {{\left|{{d\delta(J)}\over{dJ}}\right|_{J=J_0}}\over {(I-\delta(J_0))^3}}-\omega=0
\label{res1}
\end{equation}
for eq. (\ref{Al}), and
\begin{equation}
{{n'}\over {I_0'^3}}\pm \Delta -\bar{\omega}=0
\label{res2}
\end{equation}
for eq. (\ref{2f}). In eq. (\ref{res1}) $\omega$ is given and $J_0$ is fixed by our initial conditions; eq. (\ref{res1}) therefore fixes $I_0$ for any given resonance index $n$.
Around the resonance we now expand the atomic energy term in eq. (\ref{Al}) to the second order in $x=I-I_0$ and to the first order in $y=J-J_0$ and we compare the resulting expression
\begin{equation}
{1\over {(I_0-\delta(J_0))^3}}x+ {{\left|{{d\delta(J)}\over{dJ}}\right|_{J=J_0}}\over {(I-\delta(J_0))^3}}y-{3\over {2(I_0-\delta(J_0))^4}}x^2
\end{equation}
with the similar one obtained by expanding eq. (\ref{2f})
\begin{equation}
{1\over {I_0'^3}}x \pm \Delta y -{3\over {2I_0'^4}}x^2
\end{equation}
(the constant terms have been dropped).
The linear terms cancel because of the resonance conditions eqs. (\ref{res1}) and (\ref{res2}); the equality of the two $x^2$ terms instead requires 
\begin{equation}
I_0-\delta(J_0) = I'_0 
\label{act}
\end{equation}
and eq. (\ref{res2}) thus gives us the combination $\bar{\omega}\pm \Delta$, once $n'$ has been chosen. We can therefore choose $\bar{\omega}$ and $\Delta$ so that the local Hamiltonians are equivalent at {\bf two} separate resonances. If we choose $J_0$ to be the same for both resonances, the distance in $I_0$ and in $I'_0$ will be the same because of (\ref{act}) and the two resonances will overlap at the same field amplitude in the two systems.

Since we are considering ionization at low frequencies, we choose to have equivalence at the two lowest resonances: let us write $\left|{{d\delta}\over{dJ}}\right|=l+\gamma$ with $\gamma\in (0,1)$; the two lowest Alkali resonances will be those with $n=l+1$ (for $m=-1$) and $n=-l$ (for $m=+1$); for the hydrogen atom it will instead be $n'=1$. We therefore have from eq. (\ref{res1}):
\begin{eqnarray}
I^{(-)}_0= \left({{\gamma}\over{\omega}}\right)^{1/3}+\delta(J_0),  \hspace{1.in} 
I^{(+)}_0= \left({{1-\gamma}\over{\omega}}\right)^{1/3}+\delta(J_0)   
\end{eqnarray}
and finally  
\begin{eqnarray}
\omega_1= {{\omega}\over{\gamma}}, \hspace{1.0in} \hspace{1.in} 
\omega_2= {{\omega}\over{1-\gamma}} \hspace{0.8in} \\ 
F_1=F{{V_{l+1,-1} (I^{(-)}_0,J_0)}\over{V_{1}(I^{(-)}_0-\delta(J_0))}}, \hspace{1.in}
F_2=F{{V_{-l,+1} (I^{(+)}_0,J_0)}\over{V_{1}(I^{(+)}_0-\delta(J_0))}} .
\end{eqnarray}

An intuitive quantum interpretation of the last two formulas is that we equate the Rabi frequencies at each of the two resonances; the first two equations can instead be seen as equating the ratios of the single microwave frequency of the Alkali system to its two atomic transition frequencies with the ratios of the two microwave frequencies of the Hydrogen system to its atomic transition frequency.

\section{QUANTUM DELOCALIZATION BORDER}

According to Ref. \cite{cas} the threshold for quantum delocalization of a (1D) Hydrogen atom in a microwave field sum of two sine waves with {\bf incommensurable} frequencies $\omega_1$ and $\omega_2$ and amplitudes $F_1$ and $F_2$ is given by the equation 
\begin{equation}
1 \ge {{(\omega_1 \omega_2)^{5/6}}\over{1.8 (F_1 F_2)^{1/2}}}.
\end{equation}
Substituting the values obtained above, and introducing the scaled quantities $F_0=F n_0^4$, and $\omega_0=\omega n_0^3$, where $n_0=I_0$ is the quantum number of the initially populated state, this becomes
\begin{equation}
F_0 \ge {{\omega_0 ^{5/3}}\over{1.8n_0 \left({{V_{l+1,-1} (I^{(-)}_0,J_0)}\over{V_{1}(I^{(-)}_0-\delta(J_0))}} {{V_{-l,+1} (I^{(+)}_0,J_0)}\over{V_{1}(I^{(+)}_0-\delta(J_0))}} \right)^{1/2}[\gamma(1-\gamma)]^{5/6}}};
\end{equation}

The 1D hydrogenic matrix element reads $0.325 (I^{(\pm)}_0-\delta(J_0))^2$ \cite{jens}; to instead evaluate the alkali matrix elements, we apply the {\bf semiclassical} theory developed in  Refs.  \cite{oum,edm}: it allows us to speedily calculate good approximations of the full quantum matrix elements that would otherwise require lengthy calculations \cite{note1}. The formula derived in Refs.  \cite{oum,edm} for the matrix element between two states with effective quantum numbers $(\nu=n-\delta,l)$ and $(\nu'=n'-\delta',l')$ reads:
\begin{eqnarray}
\left<x\right> =1.5\nu_c^2\varepsilon\Phi\nonumber
\end{eqnarray}
where the average quantum number $\nu_c$ is again $I^{(\pm)}_0-\delta(J_0)$. For high principal quantum numbers and low angular momenta we can approximate the eccentricity with $\varepsilon \approx 1$ and $\Phi$ with $g_0(\gamma)$ tabulated in Ref. \cite{edm} so that the delocalization condition reads:
\begin{equation}
F_0 \ge {{\omega_0 ^{5/3}}\over{1.8n_0}}
{{0.217}\over{ \left[ g_0(\gamma) g_0(1-\gamma) \right]^{1/2}[\gamma(1-\gamma)]^{5/6}}}.
\label{fin}
\end{equation}
The last factor on the right side of eq. (\ref{fin}) is given  in table (1) for the $s\rightarrow p$ transition of the four alkali; its maximum variation is $10\%$. Eq. (\ref{fin}) is therefore essentially independent from $\gamma$, that is from the element considered, in accordance with the numerical results from Ref. \cite{buc}. 

Substituting the approximate value $1$ for this factor, eq. (\ref{fin}) finally becomes
\begin{equation}
F_0 \ge {{\omega_0 ^{5/3}}\over{1.8n_0}}.
\label{oho}
\end{equation}
The range of validity is $max(\gamma, 1-\gamma)<\omega_0<1$, that agrees well with the region $II$ of Ref. \cite{buc,buc1}.

Removing the esplicit dependence from $n_0$ and introducing the value $\omega=2\pi\cdot 36 \hspace{.05in} GHz= 5.47\cdot10^{-6}\hspace{.05in}a.u.$ we can cast eq. (\ref{oho}) in the form :
\begin{equation}
F_0 \ge 0.0098\omega_0 ^{4/3},
\end{equation}
which allows direct comparison with the numerical results in Ref. \cite{buc,buc1}.
Scaling instead to $\bar{\omega}_0=3.4\omega_0$ and using the value $\omega=2\pi\cdot 8.867 \hspace{.05in} GHz= 1.35\cdot10^{-6}\hspace{.05in}a.u.$ we can compare our result 
\begin{equation}
F_0 \ge 0.0012\bar{\omega}_0 ^{4/3}
\end{equation}
to the experimental ones given in Ref. \cite{arn}.

\section{CONCLUSIONS}

Both the laboratory \cite{arn} and numerical \cite{buc,buc1} results are higher than our evaluation by a factor $3.7$ which is bigger than the factor $2$ we expect from the analogous evaluation for the Hydrogen atom in a monochromatic microwave field; on the other hand it is significant that -for both different microwave frequencies and different initial principal quantum numbers- the deviation factor is always the same. In particular, the absence of a dependence of $F$ from $n_0$ which is evident for regime $II$ in the numerical results shown in Fig. 5a of Ref. \cite{buc1} is in agreement with eq. (\ref{oho}), which can be written in unscaled form as $F \ge  \omega ^{5/3}/1.8$; while it is in clear disagreement with the Hydrogen delocalization border $n_0^{-1}$ dependence \cite{chi}.

In the experimental $F_0$ Vs. $\omega_0$ plot \cite{arn} the ionization threshold values for high $\omega_0$ appear to be somehow lower than our evaluation corrected by the factor $3.7$; this is most likely due to the definition of "ionization" in the experiment as excitation above a given quantum number $n_b$: increasing $\omega_0$ by the increase of $n_0$ brings the initial conditions closer to this border thus lowering the theshold \cite{notf}.  

\section{AKNOWLEDGEMENTS}

The author wishes to thank A. Buchleiner for helpful discussions and suggestions and the Max-Planck-Institut f\"ur Physik complexer Systeme in Dresden for their ospitality when working on this paper.

\begin{table}[tbp]
\begin{eqnarray}
\hline\hline  \nonumber \\
element\hspace{0.5in} \gamma \hspace{0.3in} factor\nonumber \\
\hline  \nonumber \\
Li \hspace{0.7in} 0.35 \hspace{0.3in} 1.05\nonumber \\
Na \hspace{0.7in} 0.52 \hspace{0.3in} 0.94 \nonumber \\
K \hspace{0.7in} 0.47 \hspace{0.3in} 0.94 \nonumber \\
Rb \hspace{0.7in} 0.48 \hspace{0.3in} 0.94\nonumber \\
\hline  \nonumber
\end{eqnarray}%
\caption{The $\gamma$-dependent factor in eq. (\ref{fin}) for the $s\rightarrow p$ transition of the four alkali}
\end{table}


\begin{thebibliography}{28}
\bibitem{no1} Not all the values of $\alpha$ and $\beta$ make sense: depending on all three parameters the effective radial potential energy 
\begin{eqnarray}
V_{eff} =  {{J^2}\over{2r^2}}  - {1\over r} - {{\beta e^{-\alpha r}}\over r}\nonumber,
\end{eqnarray}
has either the single minimum we desire or two separate ones.
\bibitem{note} We assume the ratio of the two frequencies to be irrational, so that only one frequency at the time can be resonant. In effect, even for rational ratios of the two frequencies, the presence of a fractionary resonance of the second frequence does only introduce wiggles on the pendulum phase space curve.
\bibitem{note1} We e.g. checked that for the $51p \rightarrow 51d$ transition of Calcium the semiclassical procedure gives $\left<x\right>=2330 a_0$ which compares well with the full quantum calculation of the same matrix element reported by Ref. \cite{gen} ($\left<x\right>=2320 a_0$).
\end{thebibliography}
\end{document}